\documentclass{siamltex}
\usepackage{psfig}

\newcommand{\union}{{\,\cup}\,}

\newcommand{\mix}[0]{\lceil\log(n-1)\rceil+1}

\newtheorem{fact}{Fact}[section]
\newtheorem{Algorithm}{Algorithm}
\newenvironment{algorithm}{\begin{Algorithm}  \rm}{\end{Algorithm}}

\begin{document}

\title{Linear-Time Approximation Algorithms for Computing Numerical
Summation with Provably Small Errors\thanks{A preliminary version of this 
work appeared as part of 
{\em Efficient minimization of numerical summation
  errors}, in Lecture Notes in Computer Science 1443: Proceedings of the 25th
  International Colloquium on Automata, Languages, and Programming, K.~G.
  Larsen, S.~Skyum, and G.~Winskel, eds., Springer-Verlag, New York, NY, 1998,
  pp.~375--386.}}

\author{Ming-Yang Kao\thanks{Department of Computer Science, Yale
University, New Haven, CT 06520. Email: kao-ming-yang@cs.yale.edu.
Supported in part by NSF Grant CCR-9531028.}  
\and 
Jie Wang\thanks{Department of Mathematical Sciences, University of
North Carolina at Greensboro, Greensboro, NC 27412. Email:
wang@uncg.edu. Supported in part by NSF Grant CCR-9424164.}}

\maketitle

\pagestyle{myheadings}

\markboth{M. Y. Kao and J. Wang}{Minimization of Numerical Summation
Errors}

\begin{abstract} 
Given a multiset $X=\{x_1,\ldots, x_n\}$ of real numbers, the {\it
floating-point set summation} problem asks for
$S_n=x_1+\cdots+x_n$. Let $E^*_n$ denote the minimum worst-case error
over all possible orderings of evaluating $S_n$.  We prove that if $X$
has both positive and negative numbers, it is NP-hard to compute $S_n$
with the worst-case error equal to $E^*_n$.  We then give the first
known polynomial-time approximation algorithm that has a provably
small error for arbitrary $X$. Our algorithm incurs a worst-case error
at most $2(\mix)E^*_n$.\footnote{All logarithms $\log$ in this
paper are base 2.}  After $X$ is sorted, it runs in $O(n)$ time.
For the case where $X$ is either all positive or all negative, we give
another approximation algorithm with a worst-case error at most
$\lceil\log\log n\rceil E^*_n$.  Even for unsorted $X$, this algorithm runs in
$O(n)$ time.  Previously, the best linear-time approximation algorithm
had a worst-case error at most $\lceil\log n\rceil E^*_n$, while
$E^*_n$ was known to be attainable in $O(n \log n)$ time using Huffman
coding.
\end{abstract}



\section{Introduction} \label{sec_1} 
Summation of floating-point numbers is ubiquitous in numerical
analysis and has been extensively studied (for example, see
\cite{Demmel84,Goldberg90,Hig93,Knu97,Kulisch86,RS88,mik96,Hig96,Torczon91}).  
This paper focuses on the {\em floating-point set
summation} problem which, given a multiset $X=\{x_1,\ldots,x_n\}$ of
real numbers, asks for $S_n = x_1+x_2+\cdots+x_n$.  Without loss of
generality, let $x_i \not= 0$ for all $i$ throughout the paper.  Here
$X$ may contain both positive and negative numbers. For such a general
$X$, previous studies have discussed heuristic methods and obtained
statistical or empirical bounds for their errors.  We take a new
approach by designing efficient algorithms whose worst-case errors are
provably small.

Our error analysis uses the standard model of floating-point arithmetic
with unit roundoff $\alpha \ll 1$:
\[
{\rm fl}(x+y) = (x+y)(1+\delta_{xy}), \mbox{where}\ |\delta_{xy}| \leq \alpha.
\] 
Since operator $+$ is applied to two operands at a time, an ordering for
adding $X$ corresponds to a binary addition tree of $n$ leaves and $n-1$
internal nodes, where a leaf is an $x_i$ and an internal node is the sum of
its two children.  Different orderings yield different addition trees,
which may produce different computed sums $\hat{S}_n$ in floating-point
arithmetic. We aim to find an optimal ordering that minimizes the error
$E_n = |\hat{S}_n - S_n|$.  Let $I_1,\ldots,I_{n-1}$ be the internal nodes
of an addition tree $T$ over $X$. Since $\alpha$ is very small even on a
desktop computer, any product of more than one $\alpha$ is negligible in our
consideration. Using this approximation,
\[\hat{S}_n \approx S_n + \sum_{i=1}^{n-1}{I_i\delta_i}.\]
Hence, $E_n \approx |\sum_{i=1}^{n-1}I_i\delta_i| \leq
\alpha\sum_{i=1}^{n-1}|I_i|$, giving rise to the following definitions:
\begin{itemize}
\item
The {\em worst-case error} of $T$, denoted by $E(T)$, is
$\alpha\sum_{i=1}^{n-1}|I_i|$.
\item
The {\em cost} of $T$, denoted by $C(T)$, is $\sum_{i=1}^{n-1}|I_i|$.
\end{itemize}
Our task is to find a fast algorithm that constructs an addition tree
$T$ over $X$ such that $E(T)$ is small.  Since $E(T) = \alpha{\cdot}C(T)$,
minimizing $E(T)$ is equivalent to minimizing $C(T)$.  We further
adopt the following notations:
\begin{itemize}
\item 
$E^*_n$ (respectively, $C^*_n$) is the minimum worst-case error
(respectively, minimum cost) over all orderings of evaluating $S_n$.
\item 
$T_{\min}$ denotes an optimal addition tree over $X$, i.e.,
$E(T_{\min})=E^*_n$ or equivalently $C(T_{\min})=C^*_n$.
\end{itemize}
In \S\ref{sec_2}, we prove that if $X$ contains both positive and
negative numbers, it is NP-hard to compute a $T_{\min}$.  In light of
this result, we design an approximation algorithm
in~\S\ref{sec_3_general} that computes a tree $T$ with $E(T) \leq
2(\mix)E^*_n$.  
After $X$ is sorted, this algorithm takes only
$O(n)$ time.  This is the first known polynomial-time approximation
algorithm that has a provably small error for arbitrary $X$.  For the
case where $X$ is either all positive or all negative, we give another
approximation algorithm in \S\ref{sec_3_single} that computes a tree
$T$ with $E(T) \leq (1+\lceil\log\log n\rceil)E^*_n$.  This algorithm
takes only $O(n)$ time even for unsorted $X$. Previously
\cite{Hig93}, the best linear-time approximation algorithm had a
worst-case error at most $\lceil\log n\rceil E^*_n$, while $E^*_n$
was known to be attainable in $O(n \log n)$ time using Huffman coding
\cite{vanLeeuwen76}.

\section{Minimizing the worst-case error is NP-hard}\label{sec_2} 

If $X$ contains both positive and negative numbers, we prove that it
is NP-hard to find a $T_{\min}$.  We first observe the following
properties of $T_{\min}$.

\begin{lemma}\label{p:one}
Let $z$ be an internal node in $T_{\min}$ with children $z_1$ and
$z_2$, sibling $u$, and parent $r$.
\begin{enumerate}
\item \label{p:one-1}
If $z > 0$, $z_1 \geq 0$, and $z_2 > 0$, then $u \geq 0$ or $r < 0$.
\item \label{p:one-2}
If $z < 0$, $z_1 \leq 0$, and $z_2 < 0$, then $u \leq 0$ or $r > 0$.
\end{enumerate}
\end{lemma}

\begin{proof}
By symmetry. we only prove the first statement.  $C(T_r) =
|r|+|z|+C_f$, where $C_f=C(T_{z_1})+C(T_{z_2})+C(T_u)$.  Assume to the
contrary that $u < 0$ and $r \geq 0$.  Then $z \geq |u|$. We swap
$T_{z_1}$ with $T_u$. Let $z' = u+z_2$.  Now $r$ becomes the parent of
$z'$ and $z_1$.  This rearrangement of nodes does not affect the value
of node $r$, and the costs of $T_{z_1}$, $T_{z_2}$, and $T_u$ remain
unchanged.  Let $T'_r$ be the new subtree with root $r$. Let $T'$ be
the entire new tree resulted from the swapping.  Since $u$ and $z_2$
have the opposite signs, $|z'| < \max\{|u|,z_2\} \leq z$.  Hence,
$C(T'_r)= r+|z'|+C_f < r+z+C_f = C(T_r)$. Thus, $C(T') < C(T_{\min})$,
contradicting the optimality of $T_{\min}$.  This completes the proof.
\end{proof}

For the purpose of proving that finding a $T_{\min}$ is NP-hard, we
restrict all $x_i$ to nonzero integers and consider the following
optimization problem.

{\sc MINIMUM ADDITION TREE (MAT)} 

{\em Input:} A multiset $X$ of $n$ nonzero integers $x_1, \ldots,
x_n$.

{\em Output:} Some $T_{\min}$ over $X$.

The following problem is a decision version of MAT.

{\sc ADDITION TREE (AT)} 

{\em Instance:} A multiset $X$ of $n$ nonzero integers $x_1, \ldots,
x_n$, and an integer $k\geq 0$.

{\em Question:} Does there exist an addition tree $T$ over $X$ with
$C(T) \leq k$?

\begin{lemma} \label{l:2.0}
If {\rm MAT} is solvable in time polynomial in $n$, 
then {\rm AT} is also solvable in time polynomial in $n$.
\end{lemma}
\begin{proof}
Straightforward.
\end{proof}
In light of Lemma~\ref{l:2.0}, to prove that MAT is NP-hard, it
suffices to reduce the following NP-complete problem \cite{gj79} to
AT.

{\sc 3-PARTITION (3PAR) }

{\em Instance:} A multiset $B$ of $3m$ positive integers $b_1, \ldots,
b_{3m}$, and a positive integer $K$ such that $K/4 < b_i < K/2$ and
$b_1 + \cdots + b_{3m} = mK$.  

{\em Question:} Can $B$ be partitioned into $m$ disjoint sets $B_1,
\ldots, B_m$ such that for each $B_i$, $\sum_{b\in B_i}b = K$?  ($B_i$
must therefore contain exactly three elements from $B$.)

Given an instance $(B,K)$ of 3PAR, let 
\[W = 100(5m)^2K;\  a_i =
b_i+W;\  A = \{a_1, \ldots, a_{3m}\};\ L=3W+K.
\] 

\begin{lemma} \label{l:amplify}
$(A,L)$ is an instance of {\rm 3PAR}. Furthermore, it is a positive
instance if and only if $(B,K)$ also is.
\end{lemma}

\begin{proof}
Since $K/4 < b_i < K/2$, $K/4+W < a_i < K/2+W$ and thus $L/4 <
a_i < L/2$.  Next, $a_1 +a_2+\cdots + a_{3m} = 3mW+mK = mL$. This
complete the proof of the first statement.  The second statement
follows from the fact that $b_i+b_j+b_k = K$ if and only if
$a_i+a_j+a_k = L$.
\end{proof}

Write 
\[
\epsilon = \frac{1}{400(5m)^2};\ 
h = \lfloor 4\epsilon L \rfloor;\ 
H = L+h;\] 
\[
h = \beta_0 H;\ 
a_i = \left(\frac{1}{3}+\beta_i\right)\!H;\ 
a_i = \left(\frac{1}{3}+\epsilon_i\right)\!L;\  
a_M = \max\{a_i: i = 1, \ldots, 3m\}.
\]

\begin{lemma}\label{l:2.1}
\begin{enumerate}
\item 
$|\epsilon_i| < \epsilon$ for $i = 1,\ldots,3m$.
\item 
$0 < \beta_0 < 4\epsilon$, and 
$|\beta_i| < 4\epsilon$ for $i = 1,\ldots,3m$.
\item $3a_M  < H$.
\end{enumerate}
\end{lemma}

\begin{proof}

Statement 1.  Note that $b_i+W = (1/3+\epsilon_i)(3W+K)$.  Thus, $b_i
= K/3 + \epsilon_i(300(5m)^2+1)K$.  Since $K/4 < b_i < K/2$, $-1/12<
\epsilon_i(300(5m)^2+1) < 1/6$. Hence, $4(5m)^2|\epsilon_i| < 10^{-2}$, i.e., 
$|\epsilon_i| < \epsilon$.

Statement 2. Since $4{\epsilon}L > 1$, we have $\beta_0 > 0$.  
Also, since $H > L$ and $\beta_0 H = \lfloor 4\epsilon L\rfloor$,
we have $\beta_0 < 4\epsilon$. 
Next, for each $a_i$, we have 
$\beta_i = (\epsilon_i L - h/3)/(L+h)$. Then by the triangular
inequality and Statement 1, $|\beta_i| < 7\epsilon/3 < 4\epsilon$.

Statement 3. By Statement 1, $a_i < (1/3+\epsilon)L$. Thus $3a_M <
L+3\epsilon L$.  Then, since $3\epsilon L < 3K
\leq h$, $3a_M < L+h = H$.
\end{proof}

To reduce $(A,L)$ to an instance of AT, we consider a particular
multiset 
\[X = A \union \{-H, \ldots, -H\} \union \{h, \ldots, h\}\]
with $m$ copies of $-H$ and $h$ each.  Given a node $s$ in $T_{\min}$,
let $T_s$ denote the subtree rooted at $s$.  For convenience, also let
$s$ denote the value of node $s$.  Let $v(T_{\min})$ denote the value
of the root of $T_{\min}$, which is always $0$.  For brevity, we use
$\lambda$ with or without scripts to denote the sum of at most $5m$
numbers in the form of $\pm\beta_i$.  Then all nodes are in the form
of $(N/3+\lambda)H$ for some integer $N$ and some $\lambda$.  Since by
Lemma~\ref{l:2.1}, $|\lambda| \leq (5m)(4\epsilon) = (500m)^{-1}$, the
terms $N$ and $\lambda$ of each node are uniquely determined.  The
nodes in the form of $\lambda H$ are called the {\em type-$0$}
nodes. Note that $T_{\min}$ has $m$ type-0 leaves, i.e., the $m$
copies of $h$ in $X$.

\begin{lemma} \label{l:np-five}
In $T_{\min}$, type-0 nodes can only be added to type-0 nodes.
\end{lemma}

\begin{proof}
Assume to the contrary that a type-0 node $z_1$ is added to a node
$z_2$ in the form of $(\pm N/3+\lambda)H$ with $N \geq 1$.  Then
$|z_1+z_2| \geq (1/3+\lambda')H$ for some $\lambda'$. Let $z$ be the
parent of $z_1$ and $z_2$. Since $v(T_{\min}) = 0$, $z$ cannot be the
root of $T_{\min}$.  Let $u$ be the sibling of $z$. Let $r$ be the
parent of $z$ and $u$.  Let $t$ be the root of $T_{\min}$. Let $P_r$
be the path from $t$ to $r$ in $T_{\min}$.  Let $m_r$ be the number of
nodes on $P_r$.  Since $T_{\min}$ has $5m-1$ internal nodes, $m_r <
5m-1$.

We rearrange $T_{\min}$ to obtain a new tree $T'$ as follows.  First,
we replace $T_z$ with $T_{z_2}$; i.e., $r$ now has subtrees $T_{z_2}$
and $T_u$.  Let $T''$ be the remaining tree; i.e., $T''$ is $T_{\min}$
after removing $T_{z_1}$.  Next, we create $T'$ such that its root has
subtrees $T_{z_1}$ and $T''$.  This tree rearrangement eliminates the
cost $|z_1+z_2|$ from $T_r$ but may result in a new cost in the form
of $\lambda H$ on each node of $P_r$.  The total of these extra costs,
denoted by $C_\lambda$, is at most $m_r (5m)(4\epsilon)H <
(5m-1)(5m)(4\epsilon)H$. Then, $C(T') =
C(T_{\min})-|z_1+z_2|+C_\lambda \leq
C(T_{\min})-(1/3+\lambda')H+C_\lambda < C(T_{\min})+(-1/3+(5m)^2(4
\epsilon))H = C(T_{\min})+(-1/3+10^{-2})H < C(T_{\min})$,
contradicting the optimality of $T_{\min}$.  This completes the proof.
\end{proof}

\begin{lemma} \label{l:np-one}
Let $z$ be a node in $T_{\min}$. 
\begin{enumerate}
\item If $z < 0$, then $|z| \leq H$. 
\item If $z > 0$, then $z < H$.
\end{enumerate}
\end{lemma}

\begin{proof} 

Statement 1.  Assume that the statement is untrue.  Then, since all
negative leaves have values $-H$, some negative internal node $z$ has
an absolute value greater than $H$ and two negative children $z_1$ and
$z_2$.  Since $v(T_{\min}) = 0$, some $z$ has a positive sibling $u$.
We pick such a $z$ at the lowest possible level of $T_{\min}$.  Let
$r$ be the parent of $z$ and $u$. By Lemma~\ref{p:one}(\ref{p:one-2}), 
$r > 0$.  Then
$u > |z| > H$. Since all positive leaves have values less than $H$,
$u$ is an internal node with two children $u_1$ and $u_2$.  
Since $u >0$, $z<0$, and $r > 0$,
by Lemma~\ref{p:one}(\ref{p:one-1}), 
$u$ must have a positive child and a negative
child. Without loss of generality, let 
$u_1$ be positive and $u_2$ be negative.  
Then $u = u_1 - |u_2|$.
Since $z$ is at the lowest
possible level, $|u_2| \leq H$, for otherwise we could find a $z$ at a
lower level under $u_2$.  We swap $T_z$ with $T_{u_2}$.  Let $T'_r$ be
the new subtree rooted $r$. Let $u' = u_1+z$.  Since $u_2+u' = r > 0$
and $u_2 < 0$, we have $u' > 0$.  Since $|u_2| \leq H < |z|$, we have 
$u' = u_1 - |z| < u_1 - |u_2| = u$.  Let $C_f=C(T_z)+C(T_{u_1})+C(T_{u_2})$.
Then, $C(T'_r)=r+u'+C_f < r+u+C_f =C(T_r)$, which contradicts the
optimality of $T_{\min}$ because the costs of the internal nodes not
mentioned above remain unchanged.

Statement 2.  Assume that this statement is false.  Then, since all
positive leaves have values less than $H$, some internal node $z$
has a value at least $H$ as well as two positive children.  Since
$v(T_{\min}) =0$, some such $z$ has a negative sibling $u$.  By
Statement 1, $|u| \leq H$. Hence $z+u \geq 0$, contradicting Lemma
\ref{p:one}(\ref{p:one-1}).
\end{proof}

The following lemma strengthens Lemma \ref{l:np-one}.

\begin{lemma}\label{l:np-three}
\begin{enumerate}
\item \label{l:np-three-1}
Let $z$ be a node in $T_{\min}$. If $z > 0$, then $z$ is in the form
of $\lambda H$, $(1/3+\lambda)H$, or $(2/3+\lambda)H$.
\item \label{l:np-three-2}
Let $z$ be an internal node in $T_{\min}$. If $z < 0$, then $z$ is in
the form of $\lambda H$, $(-1/3+\lambda)H$, or $(-2/3+\lambda)H$.
\end{enumerate}
\end{lemma}

\begin{proof}

Statement 1.  By Lemma \ref{l:np-one}, $z < H$.  Thus, $z =
(N/3+\lambda)H$ with $0 \leq N \leq 3$.  To rule out $N=3$ by
contradiction, assume $z = (1+\lambda)H$ with $\lambda < 0$. Since by
Lemma~\ref{l:2.1} all positive leaves have values less than
$(1/3+4\epsilon)H$, $z$ is an internal node.  By Lemmas
\ref{l:np-five} and \ref{l:np-one}, $z$ has two children $z_1 =
(2/3+\lambda')$ and $z_2 = (1/3+\lambda'')$.  Since $v(T_{\min}) = 0$,
$z$ is not the root and by Lemmas \ref{l:np-five} and \ref{l:np-one},
$z$ has a negative sibling $u$.  By Lemma \ref{l:np-one}, $|u| \leq
H$.  Let $r$ be the parent of $z$ and $u$.  Then $C(T_r) =
|r|+z+C(T_{z_1})+C(T_{z_2})+C(T_u)$.  We swap $T_{z_2}$ with
$T_u$. Let $z'$ be the parent of $z_1$ and $u$.  Now $r$ is the parent
of $z'$ and $u$. Let $T'_r$ be the new subtree rooted at $r$ after the
swapping. Since $r$ remains the same, $C(T'_r) =
|r|+|z'|+C(T_{z_1})+C(T_{z_2})+C(T_u)$.  If $|u| \geq z_1$, then $|z'|
= |u|-z_1 \leq H - z_1 = (1/3-\lambda')H < z_1 < z$; otherwise, $|u| <
z_1$ and thus $|z'| = z_1-|u| < z_1 < z$.  In either case, $C(T'_r) <
C(T_r)$, contradicting the optimality of $T_{\min}$.

Statement 2.  The proof is similar to that of Statement 1. By Lemma
\ref{l:np-one}, $z = (-N/3+\lambda)H$ with $0 \leq N \leq 3$.  To rule
out $N=3$ by contradiction, assume $z=(-1+\lambda)H$ with $\lambda <
0$.  By Lemmas \ref{l:np-five} and \ref{l:np-one}, $z$ has a positive
sibling $u < H$ and two children $z_1 = (-2/3+\lambda')H$ and $z_2 =
(-1/3+\lambda'')H$.  Let $r$ be the parent of $z$ and $u$.  Then
$C(T_r) = |r|+|z|+C(T_{z_1})+C(T_{z_2})+C(T_u)$.  We swap $T_{z_2}$
with $T_u$. Let $z'$ be the parent of $z_1$ and $u$.  Now $r$ is the
parent of $z'$ and $u$. Let $T'_r$ be the new subtree rooted at $r$
after the swapping. Since $r$ is the same, $C(T'_r) =
|r|+|z'|+C(T_{z_1})+C(T_{z_2})+C(T_u)$.  If $u \geq |z_1|$, then $|z'|
= u - |z_1| < (1/3-\lambda')H < |z|$; otherwise, $u < |z_1|$ and thus
$|z'| = |z_1|-u < |z_1| < |z|$.  So $C(T'_r) < C(T_r)$, contradicting
the optimality of $T_{\min}$.
\end{proof}

The following lemma supplements
Lemma~\ref{l:np-three}(\ref{l:np-three-1}).

\begin{lemma} \label{l:np-leaf} Let $z$ be a node in $T_{\min}$. If $z
= (1/3+\lambda)H$, then $z$ is a leaf.
\end{lemma}

\begin{proof}
Assume to the contrary that $z = (1/3+\lambda)H$ is not a leaf.  By
Lemmas \ref{l:np-five} and \ref{l:np-three}, $z$ has two children $z_1
= (2/3+\lambda_1)H$ and $z_2 = (-1/3+\lambda_2)H$.  
By Lemmas \ref{l:np-five} and \ref{l:np-three},
$z_1$ has two children $z_3 = (1/3+\lambda_3)H$ and $z_4 =
(1/3+\lambda_4)H$, contradicting Lemma~\ref{p:one}(\ref{p:one-1}).
\end{proof}

The following lemma strengthens Lemma \ref{l:np-three}(\ref{l:np-three-2}).

\begin{lemma} \label{l:np-four}
Let $z$ be an internal node in $T_{\min}$. If $z < 0$, then $z$ can only
be in the form of $\lambda H$ or $(-1/3+\lambda)H$. 
\end{lemma}

\begin{proof}
To prove the lemma by contradiction, by Lemma \ref{l:np-three}, we
assume $z = (-2/3+\lambda)H$.  Let $z_1$ and $z_2$ be the two children
of $z$.  Let $u$ be the sibling of $z$; by Lemmas \ref{l:np-five} and
\ref{l:np-three}, $u = (2/3+\lambda')H$ or $(1/3+\lambda')H$.  
Let $r$ be the parent of $z$ and $u$. Then $C(T_r) =
|r|+|z|+C(T_{z_1})+C(T_{z_2})+C(T_u)$.  By Lemmas \ref{l:np-five} and
\ref{l:np-three}, there are two cases based on the values of $z_1$ and
$z_2$ with the symmetric cases omitted.

{\it Case 1}: $z_1 = (-1/3+\lambda_1)H$ and $z_2 = (-1/3+\lambda_2)H$.
Swap $T_u$ with $T_{z_2}$.  Let $z'$ be the new parent of $z_1$ and
$u$. Then $r$ is the parent of $z'$ and $u$. Let $T'_r$ be the new
subtree rooted at $r$.  Then $C(T'_r) =
|r|+|z'|+C(T_{z_1})+C(T_{z_2})+C(T_u)$.  Whether $u = (2/3+\lambda')H$
or $(1/3+\lambda')H$, we have 
$|z'| < |z|$ and thus $C(T'_r) < C(T_r)$, which
contradicts the optimality of $T_{\min}$.

{\it Case 2}: $z_1 = (1/3+\lambda_1)H$ and $z_2 = -H$.  There are two
subcases based on $u$.

{\it Case 2A}: $u = (2/3+\lambda')H$.  We swap $T_{z_1}$ with
$T_u$. Let $z'$ be the new parent of $z_2$ and $u$. Then $|z'| < |z|$.

{\it Case 2B}: $u = (1/3+\lambda')H$.  We swap $T_{z_2}$ with $T_u$.
Let $z'$ be the new parent of $z_1$ and $u$.  By Lemma
\ref{l:np-leaf}, both $z_1$ and $u$ are leaves, and thus by Lemma
\ref{l:2.1}, $2z_1+u < H$. Therefore, $|z'| = z_1+u < H-z_1 = |z|$.

Therefore, in either subcase of Case 2 the swapping results in an
addition tree over $X$ with smaller cost than $T_{\min}$, reaching a
contradiction.
\end{proof}

\begin{lemma} \label{l:np-six}
$C(T_{\min}) \geq m(H+h)$. Moreover, $C(T_{\min}) = m(H+h)$ if and only if
$(A,L)$ is a positive instance of {\rm 3PAR} .
\end{lemma}

\begin{proof} 
By Lemmas \ref{l:np-five}, \ref{l:np-three}, \ref{l:np-leaf}, and
\ref{l:np-four}, each $a_i \in A$ can only be added to some $a_j \in A$ or
to some $z_1 = (-1/3+\lambda_1)H$. In turn, $z_1$ can only be the sum of
$-H$ and some $z_2 = (2/3+\lambda_2)H$. In turn, $z_2$ is the sum of
some $a_k$ and $a_\ell \in A$.  Hence, in $T_{\min}$, $2m$ leaves in $A$
are added in pairs. The sum of each pair is then added to a leaf node
$-H$. This sum is then added to a leaf node in $A$.  This sum is a
type-0 node with value $-|\lambda'|H$, which can only be added to
another type-0 node.  Let $a_{p,1}, a_{p,2}, a_{p,3}$ be the three
leaves in $A$ associated with each $-H$ and added together as
$((a_{p,1}+a_{p,2})+(-H))+a_{p,3}$ in $T_{\min}$.  The cost of such a
subtree is $2H-(a_{p,1}+a_{p,2}+a_{p,3})$.  There are $m$ such
subtrees $R_p$. Their total cost is $2mH-\sum_{i=1}^{3m}a_i = mH+mh$.
Hence, $C(T_{\min}) \geq mH+mh$.

If $(A,L)$ is not a positive instance of {\rm 3PAR}, then for any
$T_{\min}$, there is some subtree $R_p$ with $a_{p,1}+a_{p,2}+a_{p,3}
\not= L$.  Then, the value of the root $r_i$ of $R_p$ is
$a_{p,1}+a_{p,2}+a_{p,3}-H \not= -h$.  Since $r_i$ is a type-0 node,
it can only be added to a type-0 node.  
No matter how the $m$ root values
$r_k$ and the $m$ leaves $h$ are added, some node resulting from
adding these $2m$ numbers is nonzero.  Hence, $C(T_{\min}) > mH+mh$.

If $(A,L)$ is a positive instance of {\rm 3PAR}, let
$\{a_{p,1},a_{p,2},a_{p,3}\}$ with $1 \leq p \leq m$ form a 3-set
partition of $A$; i.e., $A$ is the union of these $m$ 3-sets and for
each $p$, $a_{p,1}+a_{p,2}+a_{p,3} = L$.  Then each 3-set can be added
to one $-H$ and one $h$ as $(((a_{p,1}+a_{p,2})+(-H))+a_{p,3})+h$,
resulting in a node of value zero and contributing no extra
cost. Hence, $C(T_{\min}) = mH+mh$.  This completes the proof.
\end{proof}

\begin{theorem} \label{t:np-hard}
It is NP-hard to compute an optimal addition tree over a multiset that
contains both positive and negative numbers.
\end{theorem}
\begin{proof}
By Lemma~\ref{l:2.0}, it suffices to construct a reduction $f$ from
{\rm 3PAR} to AT. Let $f(B,K) = (X, mH+mh)$, which is polynomial-time
computable.  By Lemma~\ref{l:np-six}, $(X, mH+mh)$ is a positive
instance of AT if and only if $(A,L)$ is a positive instance of 3PAR.
Then, by Lemma~\ref{l:amplify}, $f$ is a desired reduction.
\end{proof}

\section{Approximation algorithms} \label{s2} In light of Theorem
\ref{t:np-hard}, for $X$ with both positive and negative numbers, no
polynomial-time algorithm can find a $T_{\min}$ unless P = NP
\cite{gj79}.  This motivates the consideration of approximation
algorithms.

\subsection{Linear-time approximation for general $X$} 
\label{sec_3_general}
This section assumes that $X$ contains at least one positive number
and one negative number.  We give an approximation algorithm whose
worst-case error is at most $2(\mix)E^*_n$.  If $X$ is
sorted, this algorithm takes only $O(n)$ time.

In an addition tree, a leaf is {\em critical} if its sibling is a leaf
with the opposite sign.  Note that if two leaves are siblings, then
one is critical if and only if the other is critical.  Hence, an
addition tree has an even number of critical leaves.

\begin{lemma} \label{l:critical}
Let $T$ be an addition tree over $X$.  Let $y_1, \ldots, y_{2k}$ be
its critical leaves, where $y_{2i-1}$ and $y_{2i}$ are siblings.  Let
$z_1, \ldots, z_{n-2k}$ be the noncritical leaves.  Let $\Pi =
\sum_{i=1}^{k}|y_{2i-1}+ y_{2i}|$, and $\Delta = \sum_{j=1}^{n-2k}|z_j|$.
Then $C(T) \geq (\Pi+\Delta)/2$.
\end{lemma}

\begin{proof}
Let $x$ be a leaf in $T$. There are two cases.

{\it Case 1}: $x$ is some critical leaf $y_{2i-1}$ or $y_{2i}$.  Let
$r_i$ be the parent of $y_{2i-1}$ and $y_{2i}$ in $T$ for $1 \leq i
\leq k$. Then $|r_i|=|y_{2i-1}+y_{2i}|$.

{\it Case 2}: $x$ is some noncritical leaf $z_j$.  Let $w_j$ be the
sibling of $z_j$ in $T$. Let $q_j$ be the parent of $z_j$ and
$w_j$. There are three subcases.

{\it Case 2A}: $w_j$ is also a leaf. Since $z_j$ is noncritical, $w_j$
has the same sign as $z_j$ and is also a noncritical leaf. Thus,
$|q_j| = |z_j|+|w_j|$.

{\it Case 2B}: $w_j$ is an internal node with the same sign as
$z_j$. Then $|q_j| \geq |z_j|$.

{\it Case 2C}: $w_j$ is an internal node with the opposite sign to
$z_j$. If $|w_j| \geq |z_j|$, then $|q_j|+|w_j| \geq |z_j|$; if $|w_j|
< |z_j|$, then $|q_j|+|w_j| = |z_j|$. So, we always have
$|q_j|+|w_j| \geq |z_j|$.

Observe that
\[
C(T) \geq \sum_{i=1}^k |r_i| + \frac{1}{2}\left(\sum_{z_j\
\mbox{in Case 2A}} |q_j|\right) + \sum_{z_j\ \mbox{in Case 2B}} |q_j| +
\sum_{z_j\ \mbox{in Case 2C}} |q_j|;
\]
\[
C(T) \geq \sum_{z_j\ \mbox{in Case 2C}} |w_j|. 
\]
Simplifying the sum of these two inequalities based on the case
analysis, we have $2C(T) \geq \Pi + \Delta$ as desired.
\end{proof}

In view of Lemma \ref{l:critical}, we desire to minimize $\Pi+\Delta$ over
all possible $T$.  Given $x_t, x_{t'} \in X$ with $t \neq t'$, $(x_t,
x_{t'})$ is a {\it critical pair} if $x_t$ and $x_{t'}$ have the
opposite signs. A {\it critical matching} $R$ of $X$ is a set
$\{(x_{t_{2i-1}},x_{t_{2i}}): i=1,\ldots,k\}$ of critical pairs where
the indices $t_j$ are all distinct.  For simplicity, let $y_j =
x_{t_j}$. Let $\Pi =\sum_{i=1}^k|y_{2i-1}+y_{2i}|$ and $\Delta =
\sum_{z \in X-\{y_1,\ldots,y_{2k}\}}|z|$.  If $\Pi+\Delta$ is the minimum
over all critical matchings of $X$, then $R$ is called a {\em minimum
critical matching} of $X$. Such an $R$ can be computed as follows.
Assume that $X$ consists of $\ell$ positive numbers $a_1 \leq
\cdots \leq a_\ell$ and $m$ negative numbers $-b_1 \geq \cdots \geq -b_m$.

\begin{algorithm} \label{alg:a}

\begin{enumerate}
\item If $\ell=m$, let $R = \{(a_i, -b_i): i = 1, \ldots, \ell\}$.

\item If $\ell < m$, let $R = \{(a_i, -b_{i+m-\ell}): i = 1, \ldots, \ell\}$.

\item If $\ell > m$, let $R = \{(a_{i+\ell-m},-b_i): i = 1, \ldots, m\}$.
\end{enumerate}
\end{algorithm}

\begin{lemma} \label{l:min}
If $X$ is sorted, then Algorithm \ref{alg:a} computes a minimum
critical matching $R$ of $X$ in $O(n)$ time.
\end{lemma}

\begin{proof}
By case analysis, if $a_i \leq a_j$ and $b_{i'} \leq b_{j'}$, then
$|a_i-b_{i'}|+|a_j-b_{j'}| \leq |a_i-b_{j'}|+|a_j-b_{i'}|$.  Thus, if
$\ell = m$, then pairing $a_i$ with $-b_i$ returns the minimum $\Pi+\Delta$.
For the case $\ell < m$, let $\epsilon$ be an infinitesimally small
positive number.  Let $X'$ be $X$ with additional $m-\ell$ copies of
$\epsilon$. Then, $\sum_{i=1}^\ell|a_i-b_{i+m-\ell}|
+\sum_{i=1}^{m-\ell}|\epsilon-b_i| = (\ell-m)\epsilon + \Pi + \Delta$ is the
minimum over all possible critical matchings of $X'$. Thus, $\Pi+\Delta$ is
the minimum over all possible critical matching of $X$. The case $\ell >
m$ is symmetric to the case $\ell < m$.  Since $X$ is sorted, the running
time of Algorithm \ref{alg:a} is $O(n)$.
\end{proof}

We now present an approximation algorithm to compute the summation
over $X$.

\begin{algorithm} \label{alg:b}
\begin{enumerate}
\item 
Use Algorithm \ref{alg:a} to find a minimum critical matching $R$ of
$X$. The numbers $x_i$ in the pairs of $R$ are the critical leaves in
our addition tree over $X$ and those not in the critical pairs are the
noncritical leaves.
\item 
Add each critical pair of $R$ separately.
\item 
Construct a balanced addition tree over the resulting sums of Step 2
and the noncritical leaves.
\end{enumerate}
\end{algorithm}

\begin{theorem} \label{t:approx}
Let $T$ be the addition tree over $X$ constructed by Algorithm
\ref{alg:b}.  If $X$ is sorted, then $T$ can be obtained in $O(n)$
time and $E(T) \leq 2(\mix)E(T_{\min})$.
\end{theorem}

\begin{proof}
Steps 2 and 3 of Algorithm~\ref{alg:b} both take $O(n)$ time.  By
Lemma~\ref{l:min}, Step 1 also takes $O(n)$ time and thus Algorithm
\ref{alg:b} takes $O(n)$ time.  As for the error analysis, let $T'$ be
the addition tree constructed at Step 3.  Then $C(T) = C(T')+\Pi$.
Let $h$ be the number of levels of $T'$.  Since $T'$ is a balanced
tree, $C(T') \leq (h-1)(\Pi+\Delta)$ and thus $C(T) \leq
h(\Pi+\Delta)$.  By assumption, $X$ has at least two numbers with the
opposite signs. So there are at most $n-1$ numbers to be added
pairwise at Step 3.  Thus, $h \leq \mix$. Next, by
Lemma~\ref{l:critical}, since $R$ is a minimum critical matching of
$X$, we have $C(T_{\min}) \geq (\Pi+\Delta)/2$. In summary, $E(T) \leq
2(\lceil\log(n-1)\rceil+1)E(T_{\min})$.
\end{proof}

\subsection{Improved  approximation for single-sign $X$} 
\label{sec_3_single}
This section assumes that all $x_i$ are positive; the symmetric case
where all $x_i$ are negative can be handled similarly.

Let $T$ be an addition tree over $X$.  Observe that $C(T)=
\sum_{i=1}^n x_id_i$, where $d_i$ is the number of edges on the path
from the root to the leaf $x_i$ in $T$.  Hence, finding an optimal
addition tree over $X$ is equivalent to constructing a Huffman tree to
encode $n$ characters with frequencies $x_1,
\ldots, x_n$ into binary strings
\cite{vanLeeuwen76}.

\begin{fact}\label{t:huffman}
If $X$ is unsorted $($respectively, sorted$)$, then a $T_{\min}$ over
$X$ can be constructed in $O(n\log n)$ $($respectively, $O(n)$$)$
time.
\end{fact}
\begin{proof}
If $X$ is unsorted $($respectively, sorted$)$, then a Huffman tree
over $X$ can be constructed in $O(n\log n)$
\cite{clr} $($respectively, $O(n)$ \cite{vanLeeuwen76}$)$ time.
\end{proof}

For the case where $X$ is unsorted, many applications require faster
running time than $O(n\log n)$. Previously, the best $O(n)$-time
approximation algorithm used a balanced addition tree and thus had a
worst-case error at most $\lceil\log n\rceil E^*_n$.  Here we
provide an $O(n)$-time approximation algorithm to compute the sum over
$X$ with a worst-case error at most $\lceil\log\log n\rceil
E^*_n$.  More generally, given an integer parameter $t > 0$, we wish
to find an addition tree $T$ over $X$ such that $C(T) \leq
C(T_{\min})+t\cdot |S_n|$.

\begin{algorithm} \label{alg:c}
\begin{enumerate}
\item Let $m = \lceil n/2^t \rceil$. Partition $X$ into $m$
disjoint sets $Z_1, \ldots, Z_m$ such that each $Z_i$ has
exactly $2^t$ numbers, except possibly $Z_m$, which may have less
than $2^t$ numbers.

\item For each $Z_i$, let $z_i = \max\{x:x\in Z_i\}$. Let $M = \{z_i:
1\leq i \leq m\}$. 

\item For each $Z_i$, construct a balanced addition tree $T_i$ over $Z_i$.

\item Construct a Huffman tree $H$ over $M$.

\item Construct the final addition tree $T$ over $X$ from $H$ by replacing
$z_i$ with $T_i$.
\end{enumerate}
\end{algorithm}

\begin{theorem} \label{l:approximation1}
Assume that $x_1, \ldots, x_n$ are all positive.  For any integer $t >
0$, Algorithm \ref{alg:c} computes an addition tree $T$ over $X$ in
$O(n+m\log m)$ time with $C(T) \leq C(T_{\min})+ t |S_n|$, where $m =
\lceil n/2^t\rceil$. Since $|S_n| \leq C(T_{\min})$, $E(T) \leq
(1+t)E(T_{\min})$.
\end{theorem}

\begin{proof}
For an addition tree $L$ and a node $y$ in $L$, the {\em depth} of $y$
in $L$, denoted by $d_L(y)$, is the number of edges on the path from
the root of $L$ to $y$.  Since $H$ is a Huffman tree over $M
\subseteq X$ and every $T_{\min}$ is a Huffman tree over $X$, there
exists some $T_{\min}$ such that for each $z_j$, its depth in
$T_{\min}$ is at least its depth in $H$. Furthermore, in
$T_{\min}$, the depth of each $y \in Z_i$ is at least that of $z_i$.
Therefore,
\[
\sum_{i=1}^m\sum_{x_j \in Z_i}x_j\cdot d_H(z_i) \leq C(T_{\min}).
\]
Also note that for $x_j \in Z_i$, $d_T(x_j) - d_H(z_i) \leq \log 2^t =
t$. Hence,
\begin{eqnarray*}
C(T)& = & \sum_{x_i \in X}x_i\cdot d_T(x_i) \\
& = & \sum_{i=1}^m\sum_{x_j \in Z_i}x_j\cdot d_H(z_i) +
\sum_{i=1}^m\sum_{x_j \in Z_i}x_j\cdot (d_T(x_j)-d_H(z_i)) \\
& \leq & C(T_{\min}) + t\sum_{x_i\in X}x_i
\end{eqnarray*}  
In summary, $C(T) \leq C(T_{\min})+tS_n$.  Since Step 4 takes
$O(m\log{m})$ time and the others take $O(n)$ time, the total running
time of Algorithm \ref{alg:c} is as stated.
\end{proof}

\begin{corollary}
Assume that $n \geq 4$ and all $x_1,\ldots,x_n$ are positive.  Then,
setting $t=\lfloor\log((\log$ $n)-1)\rfloor$,
Algorithm~\ref{alg:c} finds an addition tree $T$ over $X$ in $O(n)$
time with $E(T)\leq\lceil\log\log{n}\rceil E(T_{\min})$.
\end{corollary}
\begin{proof} 
Follows from Theorem \ref{l:approximation1}.
\end{proof}

\section*{Acknowledgments} 
The authors thank Don Rose, Hai Shao, Xiaobai Sun, and Steve Tate for
helpful discussions and thank the anonymous referees for their
detailed comments.

\end{document}